\documentclass[jkps,preprint,fleqn,showkeys]{revtex4-2}
\usepackage{graphicx}
\usepackage{amssymb}
\usepackage{amsmath}
\usepackage{bm}
\usepackage{color}

\begin{document}
\setcounter{page}{0}

\title[]{Energy–momentum tensor from diffeomorphism invariance in classical electrodynamics}
\author{Taeseung \surname{Choi}}
\email{tschoi@swu.ac.kr}
\affiliation{College of General Education, Seoul Women's University, Seoul
01797, Korea}
\affiliation{School of Computational Sciences, Korea Institute for Advanced Study, Seoul
02455, Korea}

\date[]{}

\begin{abstract}
We reexamine the energy–momentum tensor in classical electrodynamics from the perspective of spacetime-dependent translations, i.e., diffeomorphism invariance in flat spacetime. When energy–momentum is identified through local translations rather than constant ones, a unique, symmetric, and gauge-invariant energy–momentum tensor emerges that satisfies a genuine off shell Noether identity without invoking the equations of motion. For the free electromagnetic field, this tensor coincides with the familiar Belinfante–Rosenfeld and Bessel–Hagen expressions, but arises here directly from spacetime-dependent translation symmetry rather than from improvement procedures or compensating gauge transformations. In interacting classical electrodynamics, comprising a point charge coupled to the electromagnetic field, diffeomorphism invariance yields well-defined energy–momentum tensors for the field and the particle, while the interaction term itself generates no independent local energy–momentum tensor. Its role is instead entirely encoded in the coupled equations of motion governing energy–momentum exchange, thereby resolving ambiguities in energy–momentum localization present in canonical and improvement-based approaches.

\end{abstract}

\keywords{Energy–momentum tensor; spacetime-dependent translations; diffeomorphism invariance; Noether identities; classical electrodynamics; interacting field theory}


\maketitle

\section{Introduction}

The energy--momentum tensor is a fundamental object in classical field theory, encoding local
densities and fluxes of energy and momentum and serving as the generator of spacetime translations.
In classical electrodynamics, it is traditionally derived via Noether’s theorem applied to
constant spacetime translations of the electromagnetic action, leading to the canonical
energy--momentum tensor
\cite{Noether1918, Jackson, Landau, WeinbergQFTI, TongQFT}.
While this construction correctly reproduces conserved charges upon imposing the equations of
motion, the resulting tensor is neither symmetric nor gauge invariant and, for the electromagnetic
field, is not traceless
\cite{Jackson, Landau, WeinbergQFTI, TongQFT}.
These shortcomings are well known and have motivated various improvement procedures.

Two prominent approaches have been developed to address these issues.
The Belinfante--Rosenfeld improvement restores symmetry by adding the divergence of a superpotential
constructed from the spin current \cite{Belinfante1940, Rosenfeld1940}.
The Bessel--Hagen extension of Noether’s theorem allows symmetries that hold up to a total derivative
and, in the electromagnetic case, accompanies spacetime translations with a compensating gauge
transformation \cite{BesselHagen1921}.
For the free electromagnetic field, both methods lead to the same symmetric and gauge-invariant
energy--momentum tensor.

Despite their success, improvement-based constructions raise conceptual questions.
The local form of the energy--momentum tensor is not uniquely fixed, since total divergences can be
added without affecting conserved charges.
Moreover, conservation of the improved tensor typically holds only on shell, obscuring the
distinction between off shell Noether identities and dynamical balance equations.
These ambiguities become more pronounced in interacting theories, where improvement procedures may
yield inequivalent local tensors or rely on compensating gauge transformations whose physical status
is not always transparent.
Such issues have been discussed explicitly in interacting models
\cite{BakerCG, Singh2024}
and from a more general perspective emphasizing the structure of Noether identities and gauge
symmetries \cite{Pons2011}.

Closely related questions arise in the comparison between Noether and Hilbert definitions of the
energy--momentum tensor \cite{Hilbert1915,BakerNPB}.
By coupling a field theory to a background metric and varying the action with respect to the metric,
one obtains the Hilbert energy--momentum tensor, which is symmetric by construction.
Rather than focusing on the Noether--Hilbert correspondence itself, the present work asks whether
spacetime-dependent translation symmetry furnishes a principle by which a unique and symmetric
energy--momentum tensor arises directly, without relying on improvement procedures.

In this work, we revisit the definition of the energy--momentum tensor in classical electrodynamics
from the perspective of spacetime-dependent translations, i.e., diffeomorphism invariance in flat
spacetime.
While diffeomorphism-based constructions are known in principle, their implications for interacting
classical electrodynamics---particularly the status of the interaction term itself---have not been
systematically clarified within a unified and explicit flat-spacetime analysis.
Our aim is to provide such a clarification.

We show that identifying energy--momentum through local spacetime translations yields, for the free
electromagnetic field, a uniquely defined, symmetric, gauge-invariant, and traceless
energy--momentum tensor that satisfies a genuine off shell Noether identity without invoking the
equations of motion.
For a free relativistic point particle, the same symmetry yields only the worldline equation of
motion, reflecting the fact that the particle action is localized solely on its trajectory.
In the interacting theory, comprising the electromagnetic field, a point charge, and their
coupling, diffeomorphism invariance yields well-defined energy--momentum tensors for the field and
the particle, while the interaction Lagrangian itself gives rise to no independent local
energy--momentum tensor.
Its role is instead entirely encoded in the coupled equations of motion governing the exchange of
energy and momentum between subsystems.

This diffeomorphism-based construction provides a natural and unambiguous definition of the total
energy--momentum tensor for interacting classical electrodynamics.
It resolves ambiguities inherent in canonical and improvement-based approaches and clarifies the
relation between Noether and Hilbert energy--momentum tensors, which are shown to coincide in this
framework even in flat spacetime.

This paper is organized as follows.
In Sec.~\ref{sec:SYMCAN}, we review the canonical energy--momentum tensor and the
Belinfante--Rosenfeld and Bessel--Hagen improvement procedures.
In Sec.~\ref{sec:EMTDIFF}, we derive the energy--momentum tensors of the electromagnetic field and
the point particle from diffeomorphism invariance and analyze the interacting theory.
Section~\ref{sec:Conc} is devoted to discussion and conclusions.

%


 \section{Canonical Energy–Momentum Tensor and Its Symmetrization}
\label{sec:SYMCAN}

\subsection{Canonical energy--momentum tensor}
\label{sec:DERCAN}

For completeness, we briefly recall the canonical energy--momentum tensor for the free electromagnetic field. The action is
\begin{eqnarray}
\label{eq:FFACT}
S_F = \int d^4x\, \mathcal{L}_F(x),
\qquad
\mathcal{L}_F(x) = -\frac{1}{4} F^{\mu\nu}(x)F_{\mu\nu}(x),
\end{eqnarray}
where \( F_{\mu\nu} = \partial_\mu A_\nu - \partial_\nu A_\mu \), and repeated indices are summed over.

We consider an infinitesimal constant spacetime translation with constant parameter $a^\mu$,
\begin{eqnarray}
x'^\mu = x^\mu + a^\mu ,
\end{eqnarray}
under which the gauge potential varies as
\begin{eqnarray}
\label{eq:AFVAR}
\delta A_\mu(x)
\equiv A'_\mu(x) - A_\mu(x)
= - a^\alpha \partial_\alpha A_\mu(x),
\end{eqnarray}
where \(x\) denotes the spacetime point \(x^\mu\).

The variation of the action induced by this transformation becomes
\begin{eqnarray}
\label{eq:VAGF}
\delta S_F
= \int d^4x
\left[
E^\mu \delta A_\mu
+ \partial_\nu
\left(
\frac{\partial \mathcal{L}_F}{\partial (\partial_\nu A_\mu)} \,
\delta A_\mu
\right)
\right],
\end{eqnarray}
where \(E^\mu\) denotes the Euler--Lagrange expression; the equations of motion are satisfied when \(E^\mu=0\).

On the other hand, since the Lagrangian density is a scalar,
\(\mathcal{L}'_F(x')=\mathcal{L}_F(x)\),
the variation of the action can equivalently be written as
\begin{eqnarray}
\label{eq:VALG}
\delta S_F
= \int d^4x\,\delta\mathcal{L}_F
= -a^\nu \int d^4x\,\partial_\nu \mathcal{L}_F .
\end{eqnarray}

Equating the two expressions for \(\delta S_F\) in
Eqs.~(\ref{eq:VAGF}) and (\ref{eq:VALG}), and imposing the equations of motion, one obtains the continuity equation for the canonical energy--momentum tensor,
\begin{eqnarray}
\partial_\mu T^{\mu\nu}_C = 0 ,
\end{eqnarray}
with
\begin{eqnarray}
\label{eq:CANEMT}
T^{\mu\nu}_C
= \frac{\partial \mathcal{L}_F}{\partial(\partial_\mu A_\alpha)}
\,\partial^\nu A_\alpha
- \eta^{\mu\nu}\mathcal{L}_F
= -F^{\mu\alpha}\partial^\nu A_\alpha
+ \frac{1}{4}\eta^{\mu\nu}F^{\alpha\beta}F_{\alpha\beta}.
\end{eqnarray}
Here \(\eta^{\mu\nu}=\mathrm{diag}(+,-,-,-)\) is the Minkowski metric. Here and in the following, the term "canonical energy--momentum tensor" refers to the
Noether tensor obtained from constant spacetime translations.

Although conserved on shell, the canonical energy--momentum tensor $T^{\mu\nu}_C$ is gauge
dependent, non-symmetric, and not traceless, motivating the symmetrization procedures
discussed below.

We emphasize that in this derivation the conservation law is obtained by equating two
equivalent evaluations of the same variation of the action:
one arising from the functional variation of the gauge field,
and the other from the scalar-density character of the Lagrangian under spacetime translations.
While fully equivalent to the standard derivation for constant translations,
this formulation streamlines the use of symmetry principles
and proves particularly transparent in the extension to spacetime-dependent
translations developed below. In the usual presentation, invariance of the action is imposed directly through
$\delta S_F=0$, while the scalar-density property of the Lagrangian is invoked separately
to account for the coordinate transformation; here these ingredients are combined
into a single unified variational identity.

\subsection{Symmetrization procedures}

As discussed above, the canonical energy--momentum tensor obtained from constant spacetime
translations is conserved on shell but is neither symmetric nor gauge invariant, and it is
not traceless for the electromagnetic field.
These deficiencies motivate the introduction of symmetrization procedures that modify the
local form of the energy--momentum tensor while preserving the associated conserved charges.

A widely used approach to symmetrizing the canonical energy--momentum tensor is the
Belinfante--Rosenfeld (BR) improvement, which adds the divergence of a superpotential
constructed from the spin current,
\begin{eqnarray}
T^{\mu\nu}_{\mathrm{BR}}
= T^{\mu\nu}_C + \partial_\alpha \chi^{\alpha\mu\nu},
\qquad
\chi^{\alpha\mu\nu} = -\chi^{\mu\alpha\nu}.
\end{eqnarray}
For the electromagnetic field, a convenient choice is
\(\chi^{\alpha\mu\nu} = -F^{\alpha\mu} A^\nu\).
Using the free Maxwell equations, this leads to the familiar symmetric and gauge-invariant
energy--momentum tensor \cite{Belinfante1940, Rosenfeld1940},
\begin{eqnarray}
\label{eq:SYMT}
T^{\mu\nu}_{\mathrm{Sym}}
= -F^{\mu\alpha}F^{\nu}{}_{\alpha}
+ \frac{1}{4}\eta^{\mu\nu}F^{\alpha\beta}F_{\alpha\beta}.
\end{eqnarray}
Although the BR improvement itself is defined off shell, the equations of motion are required
to eliminate explicit gauge-potential dependence and render the final expression manifestly
gauge invariant.

An alternative construction was proposed by Bessel-Hagen, who extended Noether’s
theorem to symmetries under which the Lagrangian changes by a total derivative.
In this approach, a constant spacetime translation is accompanied by a compensating
gauge transformation with parameter $\Lambda = a^\alpha A_\alpha$, leading to the
gauge-covariant variation \cite{BesselHagen1921}
\begin{eqnarray}
\label{eq:BHFVGF}
\delta A_\mu(x)
= -a^\alpha \partial_\alpha A_\mu(x) + \partial_\mu \Lambda(x)
= -a^\alpha F_{\alpha\mu}.
\end{eqnarray}
This construction yields the same symmetric energy--momentum tensor as the
Belinfante--Rosenfeld method in the free-field case.

While both procedures are algebraically consistent and lead to identical results
for free electrodynamics, neither uniquely fixes the local form of the energy--momentum
tensor. The freedom to add total divergences or to introduce compensating gauge
transformations leaves the conserved charges unchanged but introduces an intrinsic
ambiguity in energy--momentum localization. This limitation becomes especially
apparent in interacting theories and motivates the diffeomorphism-based construction
developed in the following section.

\subsection{Interacting case}

We now consider the interaction between a relativistic point particle and the
electromagnetic field, described by the action
\begin{eqnarray}
\label{eq:FCPenergy--momentumFIT}
S_I
= \int d^4x\,\mathcal{L}_I
= -e \int d^4x\, A_\mu(x) J^\mu(x),
\qquad
J^\mu(x)
= \int dt\,\dot X^\mu(t)\,\delta^{(4)}(x-X(t)),
\end{eqnarray}
where $t$ is an arbitrary worldline parameter and $X^\mu(t)$ denotes the particle
trajectory.  Equivalently, the interaction action may be written in worldline
form as
\begin{eqnarray}
S_I
= -e \int dt\, A_\mu(X(t)) \dot X^\mu(t).
\end{eqnarray}

The variation of $S_I$ takes the form
\begin{eqnarray}
\label{eq:VARIA}
\delta S_I
= -e \int dt \Big(
\dot X^\mu\,\delta_{\mathrm{field}} A_\mu(X)
+ \dot X^\mu\,\delta_X A_\mu(X)
+ A_\mu(X)\,\delta \dot X^\mu
\Big),
\end{eqnarray}
where $\delta_{\mathrm{field}}A_\mu(X)\equiv A'_\mu(X)-A_\mu(X)$ denotes the
variation of the gauge field at fixed spacetime argument, while
\begin{eqnarray}
\delta_X A_\mu(X)
\equiv A_\mu(X+\delta X)-A_\mu(X)
= \delta X^\alpha \partial_\alpha A_\mu(X)
\end{eqnarray}
is induced by the displacement of the worldline.

The first term in Eq.~(\ref{eq:VARIA}) yields the source term in Maxwell’s
equations after rewriting the worldline current in spacetime form,
\begin{eqnarray}
\label{eq:ITMXWEQ}
\partial_\mu F^{\mu\nu} = e J^\nu(x).
\end{eqnarray}
As a result, the bulk term in the symmetry variation of the free field action does
not vanish by itself in the interacting theory. Instead, upon imposing the
interacting Maxwell equations~(\ref{eq:ITMXWEQ}), it becomes equal and opposite
to the bulk contribution arising from the variation of the interaction action
with respect to the gauge field—namely, the first term in
Eq.~(\ref{eq:VARIA})—so that the two cancel on-shell.

The last term in Eq.~(\ref{eq:VARIA}) can be written as
\begin{eqnarray}
-e\int dt\, A_\mu(X)\,\delta \dot{X}^\mu
= \text{surface term}
+ e \int dt\, \delta X^\mu \frac{d}{dt}A_\mu(X).
\end{eqnarray}
We now specialize to spacetime translation symmetry, for which the variations of
the worldline and the gauge field are correlated.  Under a constant spacetime
translation,
\begin{eqnarray}
\delta X^\mu(t) = a^\mu, \qquad
\delta_X A_\mu(X)=a^\alpha \partial_\alpha A_\mu(X),
\end{eqnarray}
the remaining two terms in Eq.~(\ref{eq:VARIA}) become
\begin{eqnarray}
-e\int dt \Big(
\dot X^\mu\,\delta_X A_\mu(X)
+ A_\mu(X)\,\delta \dot X^\mu
\Big)
= -e a^\alpha \int dt\, \dot{X}^\mu F_{\alpha\mu}(X(t)),
\end{eqnarray}
neglecting total derivatives under standard boundary conditions. This is manifestly gauge invariant and, together with the variation of the free
particle action, yields the Lorentz-force equation through the worldline
Euler–Lagrange equations,
\begin{eqnarray}
m\frac{d}{dt} \dot{X}^\mu
= eF^{\mu}{}_{\alpha} \dot{X}^\alpha.
\end{eqnarray}
The surface term corresponds to a total derivative in the worldline Lagrangian
and therefore has no effect on the equations of motion. As in the
Bessel–Hagen construction, such terms do not contribute to locally conserved
momentum currents.

The interaction action does not define an independent local contribution to the
field energy--momentum tensor. Instead, in the presence of the source current
$J^\mu$, it governs the local balance of energy and momentum exchanged between
the electromagnetic field and the particle through the equations of motion.

In the Bessel--Hagen construction, the spacetime translation of the gauge field
is supplemented by a compensating gauge transformation with parameter
$\Lambda = a^\alpha A_\alpha$.  This modifies the variation of the interaction
action by a total derivative,
\begin{eqnarray}
\delta S_I
= -e \int dt\, \dot X^\mu
\partial_\mu \!\left( a^\alpha A_\alpha(X(t)) \right)
= -e \int dt\, \frac{d}{dt}
\!\left( a^\alpha A_\alpha(X(t)) \right).
\end{eqnarray}
Thus, in the interacting case, the Bessel--Hagen construction yields invariance
of the action only up to a boundary term, rather than a unique and
gauge-invariant local energy--momentum tensor for the electromagnetic field
alone.

For completeness, we note that the particle admits a local energy–momentum tensor
whose divergence equals the Lorentz-force density. Together with the corresponding
balance equation for the electromagnetic field, this implies local conservation of
the total energy–momentum tensor on shell. A fully off-shell and unambiguous
derivation is provided in Sec.~III using diffeomorphism invariance.

Nevertheless, while the symmetric energy--momentum tensor can be obtained for the free
electromagnetic field via both the Belinfante--Rosenfeld and Bessel--Hagen
constructions, neither approach yields a unique, gauge-invariant, and
unambiguous local definition of the field energy--momentum tensor in the
presence of interactions. This motivates the diffeomorphism-based construction
developed in the following section, which resolves these issues at the level of
spacetime-dependent symmetry.

\section{Energy--momentum tensor from diffeomorphism invariance}
\label{sec:EMTDIFF}

In this section we derive the energy--momentum tensor from invariance under infinitesimal
spacetime-dependent translations, i.e.\ diffeomorphisms, rather than from constant translations
as in the canonical Noether construction. Here and in the following, diffeomorphisms are understood as spacetime-dependent
translations of spacetime points in flat Minkowski spacetime; fields transform
according to their tensorial character under these transformations, without
introducing curvature or dynamical gravity. An infinitesimal diffeomorphism is defined by
\begin{eqnarray}
\label{eq:DFFCT}
x'^{\mu} = x^\mu + \epsilon^\mu(x),
\end{eqnarray}
where $\epsilon^\mu(x)$ is an arbitrary smooth vector field.

\subsection{Free electromagnetic field}
\label{sec:FRFCDF}

We first consider the free electromagnetic (Maxwell) action in Eq.~(\ref{eq:FFACT}),
\begin{eqnarray}
S_F = \int d^4x\,\mathcal{L}_F,
\qquad
\mathcal{L}_F = -\frac14 F^{\mu\nu}F_{\mu\nu}.
\end{eqnarray}

In the following we work entirely in the passive viewpoint.
Under the coordinate transformation (\ref{eq:DFFCT}), the gauge potential transforms as a one-form,
\begin{eqnarray}
A'_\mu(x')\,dx'^\mu = A_\mu(x)\,dx^\mu,
\end{eqnarray}
which implies
\begin{eqnarray}
A'_\mu(x') = \frac{\partial x^\nu}{\partial x'^\mu} A_\nu(x).
\end{eqnarray}

To compare $S'_F$ with $S_F$, we rewrite the transformed action $S'_F$ by relabeling the dummy integration variable
$x'\to x$. Expressed at the same spacetime point $x$, the induced variation of the gauge field is
\begin{eqnarray}
\label{eq:VGFDF}
\delta A_\mu(x)
\equiv A'_\mu(x)-A_\mu(x)
= -\epsilon^\alpha\partial_\alpha A_\mu
- (\partial_\mu\epsilon^\alpha)A_\alpha .
\end{eqnarray}

The variation of the action can be evaluated in two equivalent ways as in Sec. \ref{sec:SYMCAN}:
(i) by varying the fields at fixed spacetime point using $\delta A_\mu(x)$, and 
(ii) by using the scalar nature of $\mathcal L_F$ together with the Jacobian of the coordinate transformation,
$d^4x' = J\,d^4x$ with $J=\det(\partial x'/\partial x)$.
Equating the two expressions for $\delta S_F=S'_F-S_F$, one finds 
\begin{eqnarray}
\label{eq:VRFDF}
 -\int d^4x\,(\partial_\mu\epsilon_\nu)
\left(
- F^{\mu\alpha}F^\nu{}_{\alpha}
+ \frac14 \eta^{\mu\nu} F^{\alpha\beta}F_{\alpha\beta}
\right)=0,
\end{eqnarray}
using the standard variation of the Maxwell action and integrating by parts, without invoking the equations of motion.

Since $\epsilon_\nu(x)$ is an arbitrary smooth spacetime function, invariance of the action implies the off shell Noether identity
\begin{eqnarray}
\label{eq:CONETDFF}
\partial_\mu T_D^{\mu\nu} = 0,
\end{eqnarray}
with the energy--momentum tensor defined by
\begin{eqnarray}
T_D^{\mu\nu}
= - F^{\mu\alpha}F^\nu{}_{\alpha}
+ \frac{1}{4} \eta^{\mu\nu} F^{\alpha\beta}F_{\alpha\beta}.
\end{eqnarray}

The tensor $T_D^{\mu\nu}$ coincides with the symmetric, gauge-invariant energy--momentum tensor $T_{\mathrm{Sym}}^{\mu\nu}$ 
obtained by the Belinfante--Rosenfeld and Bessel--Hagen constructions in the free-field case.
Importantly, the identity $\partial_\mu T_D^{\mu\nu}=0$ follows solely from diffeomorphism invariance and holds
off shell, without invoking the Maxwell equations.

\subsection{Free relativistic particle}

We next consider a free relativistic point particle of mass $m$, whose worldline action is
\begin{eqnarray}
S_p = -m \int dt\,\sqrt{\dot X^2}
= \int d^4x\,\mathcal L_p(x),
\qquad
\mathcal L_p(x)
= -m\int dt\,\sqrt{\dot X^2}\,\delta^{(4)}(x-X(t)),
\end{eqnarray}
where $\dot X^2 = \eta_{\mu\nu}\dot X^\mu\dot X^\nu$ and $t$ is an arbitrary worldline parameter.

Under an infinitesimal diffeomorphism~(\ref{eq:DFFCT}), the worldline transforms as
\begin{eqnarray}
X'^\mu(t)=X^\mu(t)+\epsilon^\mu(X(t)).
\end{eqnarray}
The variation of the particle action may be written as
\begin{eqnarray}
\delta S_p
= \int d^4x\,\bigl(\mathcal L'_p(x)-\mathcal L_p(x)\bigr).
\end{eqnarray}
Evaluating $\mathcal L'_p(x)$ by expanding the variations of $\dot X^\mu$ and of the delta
distribution yields
\begin{eqnarray}
\delta S_p
= -m\int d^4x\,dt
\left[
\frac{\dot X_\mu}{\sqrt{\dot X^2}}
\bigl(\dot X^\alpha\partial_\alpha\epsilon^\mu(X)\bigr)\delta^{(4)}(x-X)
- \sqrt{\dot X^2}\,\epsilon^\alpha(X)\partial_\alpha\delta^{(4)}(x-X)
\right].
\end{eqnarray}
The second term is a spacetime total divergence,
\begin{eqnarray}
- \int d^4x\,\partial_\alpha\!\left(\epsilon^\alpha(x)\mathcal L_p(x)\right),
\end{eqnarray}
which is precisely the contribution obtained by treating $\mathcal L_p$ as a scalar density under
diffeomorphisms. 

As in the free-field case, we equate two equivalent expressions for the variation
of the action. In this comparison, spacetime total divergences play no role in
the final variational identity.

The remaining term can be written as
\begin{eqnarray}
\label{eq:NIDFP}
-\int d^4x\,(\partial_\alpha \epsilon_\mu)\,T_p^{\alpha\mu}(x)=0,
\end{eqnarray}
where we have introduced the kinematical energy--momentum tensor
\begin{eqnarray}
T_p^{\mu\nu}(x)
= m\int dt\,\frac{\dot X^\mu(t)\dot X^\nu(t)}{\sqrt{\dot X^2(t)}}\,
\delta^{(4)}\!\bigl(x-X(t)\bigr).
\end{eqnarray}

At first sight, one might be tempted to infer an off shell spacetime Noether identity
$\partial_\mu T_p^{\mu\nu}=0$ by arguing that the diffeomorphism parameter
$\epsilon_\mu(x)$ is arbitrary.
This inference, however, is incorrect.
Although the above relation follows algebraically from equating the two variations of the
action, the particle action is localized on the worldline. As a consequence, the diffeomorphism parameter enters the variation only through its values and derivatives evaluated at the particle position $x=X(t)$. 

Accordingly, Eq.~(\ref{eq:NIDFP}) does not represent a spacetime identity
valid pointwise.
Instead, it reduces to a relation defined along the worldline, which is
equivalent to the particle equation of motion. Indeed, integrating by parts and assuming vanishing surface terms yields
\begin{eqnarray}
\int d^4x\,\epsilon_\nu(x)\,\partial_\mu T_p^{\mu\nu}(x)
= \int dt\,\epsilon_\nu\!\bigl(X(t)\bigr)\,\frac{d p^\nu}{dt},
\qquad
p^\mu = m\frac{\dot X^\mu}{\sqrt{\dot X^2}} .
\end{eqnarray}
Since $\epsilon_\nu(X(t))$ is arbitrary only through its values evaluated on the worldline, diffeomorphism
invariance implies the worldline identity
\begin{eqnarray}
\frac{dp^\mu}{dt}=0,
\end{eqnarray}
which is precisely the equation of motion of a free relativistic particle.

Thus, for a point particle, diffeomorphism symmetry does not lead to an
off shell spacetime conservation law.
Instead, it reproduces the particle dynamics.
Only for the total system, including both fields and particles, does
diffeomorphism invariance yield an off shell spacetime identity valid for
arbitrary spacetime-dependent $\epsilon^\mu(x)$, leading to
$\partial_\mu T_T^{\mu\nu}=0$ for the total energy--momentum tensor
$T_T^{\mu\nu}=T^{\mu\nu}_D+T^{\mu\nu}_p$.

\subsection{Interacting case under diffeomorphism}

We now examine the interaction action under an infinitesimal diffeomorphism
\( x^\mu \to x^\mu + \epsilon^\mu(x) \).
To investigate the variation of the interaction action $S_I$ under such a
transformation, it is convenient to consider $S_I$ in the spacetime form
introduced in Eq.~(\ref{eq:FCPenergy--momentumFIT}),
\begin{eqnarray}
S_I
= \int d^4x\,\mathcal{L}_I
= -e \int d^4x\,dt\, A_\mu(x)\,\dot{X}^\mu(t)\,
\delta^{(4)}(x-X(t)).
\end{eqnarray}

After relabeling the dummy integration variable $x' \to x$, the transformed
action becomes
\begin{eqnarray}
S'_I
= \int d^4x\,dt\, A'_\mu(x)\,\dot{X'}^\mu(t)\,
\delta^{(4)}(x-X'(t)).
\end{eqnarray}

Under the infinitesimal diffeomorphism, the gauge field and the worldline
transform as
\begin{eqnarray}
\delta A_\mu
= -\epsilon^\alpha \partial_\alpha A_\mu
  - (\partial_\mu \epsilon^\alpha) A_\alpha,
\qquad
\delta X^\mu(t) = \epsilon^\mu(X(t)).
\end{eqnarray}
Since the delta function depends on the spacetime coordinate through its argument
$x-X(t)$, its variation under the diffeomorphism arises from the transformation
of the worldline and is given by
\begin{eqnarray}
\delta\!\left[\delta^{(4)}(x-X(t))\right]
 =\delta^{(4)}(x-X'(t))-\delta^{(4)}(x-X(t))
= -\partial_\alpha\!\left(
\epsilon^\alpha \delta^{(4)}(x-X(t))
\right).
\end{eqnarray}

Using these variations, the change of the interaction action can be written as
\begin{eqnarray}
\delta S_I
= -e \int d^4x\,dt\,
\partial_\alpha\!\left(
\epsilon^\alpha A_\mu \dot X^\mu
\delta^{(4)}(x-X(t))
\right).
\end{eqnarray}
Thus, the variation reduces entirely to a spacetime total divergence.
No term proportional to $\partial_\mu \epsilon_\nu$ appears, and therefore the
interaction action does not generate an independent Noether identity associated
with spacetime translations.

This result coincides with the equivalent evaluation
\(\delta S_I = \int d^4x \bigl(\mathcal L'_I - \mathcal L_I\bigr)\),
based on the scalar-density nature of $\mathcal L_I$ under the same
diffeomorphism. Consequently, the interaction Lagrangian does not give rise to
a local energy--momentum tensor. Its physical role is instead encoded entirely
in the coupled equations of motion for the field and the particle, which govern
the exchange of energy and momentum between them.

\subsection{Total energy--momentum tensor}

We are now ready to consider the total action of an electromagnetic field interacting with a
charged particle,
\begin{eqnarray}
S_T = S_F + S_p + S_I .
\end{eqnarray}
Under the infinitesimal diffeomorphism $x'^\mu=x^\mu+\epsilon^\mu(x)$, the total variation of the action
becomes
\begin{eqnarray}
\label{eq:TVACTDF}
\delta S_T
= -\int d^4x\,(\partial_\mu\epsilon_\nu)
\bigl(T_D^{\mu\nu} + T_p^{\mu\nu}\bigr)
=0,
\end{eqnarray}
where we have used Eqs.~(\ref{eq:VRFDF}), (\ref{eq:NIDFP}), and the fact that $\delta S_I$ reduces to a spacetime total divergence under
diffeomorphisms.

Since the diffeomorphism parameter $\epsilon^\mu(x)$ is now probed at arbitrary spacetime points
through the field contribution $T_D^{\mu\nu}$, Eq.~(\ref{eq:TVACTDF}) implies a genuine spacetime
Noether identity,
\begin{eqnarray}
\partial_\mu T_T^{\mu\nu} = 0,
\qquad
T_T^{\mu\nu} \equiv T_D^{\mu\nu}+T_p^{\mu\nu}.
\end{eqnarray}
This should be contrasted with the free particle case, where diffeomorphism
invariance yields only the worldline equation of motion, since the action is
supported solely on the particle trajectory.

Explicitly, the total symmetric energy--momentum tensor is
\begin{eqnarray}
T_T^{\mu\nu}
=
- F^{\mu\alpha}F^\nu{}_{\alpha}
+ \frac{1}{4}\eta^{\mu\nu}F^{\alpha\beta}F_{\alpha\beta}
+ m\int dt\,\frac{\dot X^\mu\dot X^\nu}{\sqrt{\dot X^2}}\,
\delta^{(4)}(x-X(t)).
\end{eqnarray}
This tensor is obtained directly from diffeomorphism invariance, without invoking
symmetrization procedures such as the Belinfante--Rosenfeld or Bessel--Hagen constructions.

While the Bessel--Hagen method also reproduces the same symmetric total energy--momentum tensor,
it relies on a specific compensating gauge transformation whose physical motivation is not manifest.
By contrast, the diffeomorphism-based construction follows directly from spacetime symmetry and
applies uniformly to both the field and particle sectors.

Moreover, diffeomorphism invariance may equivalently be viewed as arising from variations of the
background metric. From this perspective, the total energy--momentum tensor coincides with the
Hilbert definition,
\begin{eqnarray}
T_T^{\mu\nu}
=
-\frac{2}{\sqrt{-g}}
\left.\frac{\delta S_T}{\delta g_{\mu\nu}}\right|_{g_{\mu\nu}=\eta_{\mu\nu}} .
\end{eqnarray}
This agreement should be regarded as a consistency check rather than an assumption:
the present derivation does not invoke coupling to gravity or curved spacetime, but
follows purely from spacetime-dependent translation symmetry in flat spacetime.

\section{Conclusion and Discussion}
\label{sec:Conc}

In this work we have revisited the definition of the energy--momentum tensor in
classical electrodynamics from the perspective of spacetime-dependent
translations, i.e., diffeomorphism symmetry in flat spacetime. We have shown that
when energy--momentum is identified through invariance under local spacetime
translations rather than constant ones, the resulting energy--momentum tensor
for the free electromagnetic field is symmetric, gauge-invariant, and traceless.
Importantly, the associated continuity equation follows as an off-shell Noether
identity, without invoking the equations of motion.

In the free-field case, the energy--momentum tensor obtained from diffeomorphism
invariance coincides with the familiar symmetric tensor derived via the
Belinfante--Rosenfeld and Bessel--Hagen constructions. Our analysis clarifies,
however, that this agreement holds at the level of the final expression rather
than at the level of the underlying symmetry principle. In the
improvement-based approaches, symmetry in the tensor indices and gauge
invariance are achieved only after introducing specific improvement terms or
compensating gauge transformations associated with constant spacetime
translations, and the corresponding conservation laws hold only on shell. By
contrast, in the diffeomorphism-based construction these properties arise
directly from spacetime-dependent translation symmetry and are realized off
shell.

For interacting classical electrodynamics, comprising a relativistic point
particle, the electromagnetic field, and their interaction, we find a clear
structural separation. Both the field and the particle admit well-defined
energy--momentum tensors derived from diffeomorphism invariance, while the
interaction Lagrangian itself yields no independent local energy--momentum
tensor, as its variation under diffeomorphisms reduces to a spacetime total
divergence and contains no terms proportional to $\partial_\mu\epsilon_\nu$.
The role of the interaction is instead fully encoded in the coupled equations of
motion, which govern the exchange of energy and momentum between subsystems.
This perspective resolves ambiguities present in canonical and
improvement-based approaches, where it is often unclear whether interaction
effects modify the field energy--momentum tensor or merely appear through
on-shell balance relations.

Finally, we emphasize that the present analysis is carried out entirely within
flat spacetime and does not rely on coupling to gravity or curved geometry.
Nevertheless, it demonstrates that diffeomorphism invariance—understood here as
spacetime-dependent translations in Minkowski space—provides a natural and
geometrically well-defined framework for constructing energy--momentum tensors
in classical field theory. In particular, the resulting total energy--momentum
tensor coincides with the Hilbert definition obtained from metric variation,
thereby placing the analysis on a geometric footing even in flat spacetime. This
framework provides a coherent basis for understanding energy--momentum
localization and exchange in interacting classical systems and offers a natural
starting point for further developments.

%


\begin{acknowledgments}

 This work was supported by a research grant from Seoul Women’s University(2025-0047).

\end{acknowledgments}

%

\end{document}